\documentstyle[aps,prd,floats,psfig]{revtex}

\preprint{DAMTP-2000-88}
\date{25 August 2000}
\tighten

\begin{document}
\draft
\newcommand\lsim{\mathrel{\rlap{\lower4pt\hbox{\hskip1pt$\sim$}}
    \raise1pt\hbox{$<$}}}
\newcommand\gsim{\mathrel{\rlap{\lower4pt\hbox{\hskip1pt$\sim$}}
    \raise1pt\hbox{$>$}}}


\title{Looking for a varying $\alpha$ in the Cosmic Microwave Background}

\author{P. P. Avelino${}^{1,2}$\thanks{
Electronic address: pedro\,@\,astro.up.pt},
C. J. A. P. Martins${}^{3,1}$\thanks{
Electronic address: C.J.A.P.Martins\,@\,damtp.cam.ac.uk},
G. Rocha${}^{1,4}$\thanks{Electronic address: graca\,@\,astro.up.pt}
and P. Viana${}^{1,5}$\thanks{
Electronic address: viana\,@\,astro.up.pt}
}

\address{${}^1$ Centro de Astrof\'{\i}sica, Universidade do Porto,\\
Rua das Estrelas s/n, 4150-762 Porto, Portugal}

\address{${}^2$ Dep. de F{\' \i}sica da Faculdade de Ci\^encias da 
Univ. do Porto,\\
Rua do Campo Alegre 687, 4169-007 Porto, Portugal}

\address{${}^3$ Department of Applied Mathematics and Theoretical Physics\\
Centre for Mathematical Sciences, University of Cambridge\\
Wilberforce Road, Cambridge CB3 0WA, U.K.}

\address{${}^4$Department of Physics, University of Oxford,\\ 
Nuclear \& Astrophysics Laboratory, Keble Road, Oxford OX1 3RH, U.K.\\
}

\address{${}^5$ Dep. de Matem\'atica Aplicada da Faculdade de
Ci\^encias da  Univ. do Porto,\\
Rua das Taipas 135 , 4050  Porto, Portugal}

\maketitle
\begin{abstract}

{We perform a likelihood analysis of the recently released BOOMERanG and
MAXIMA data, allowing for the possibility of a time-varying fine-structure
constant. We find that in general this data prefers a value of $\alpha$ that
was smaller in the past (which is in agreement with measurements of $\alpha$
from quasar observations). However, there are some interesting degeneracies
in the problem which imply that strong statements about $\alpha$ can not
be made using this method until independent accurate determinations of
$\Omega_b h^2$ and $H_0$ are available.

We also show that a preferred lower value of $\alpha$
comes mainly from the
data points around the first Doppler peak, whereas the main effect
of the high-$\ell$ data points is to increase the preferred value
for $\Omega_b h^2$ (while also tightening the constraints on $\Omega_0$
and $H_0$).
We comment on some implications of our results.}

\end{abstract} 

\section{Introduction}
\label{intro} 

There has been a recent growth of interest in theories where some of
the usual constants of nature are actually time- and/or space-varying
quantities. Most notably, the possibility of a time-varying fine-structure
$\alpha$, has been the subject of a considerable amount of work, both at
the theoretical and experimental/observational level.

From the theoretical point of view, the motivation comes from
the recent work on the higher-dimensional theories \cite{polc},
which are thought to be required to provide a consistent
unification of the know fundamental interactions. In such theories
the `effective' three-dimensional constants are typically
related to the `true' higher-dimensional constants via the radii of the
(compact) extra dimensions \cite{banks}. On the other hand, these
radii often have a non-trivial evolution, naturally leading to the
expectation of time (or even space) variations of the `effective' coupling
constants we can measure \cite{chodos,wu,kiritsis}.

There are a number of different ways in which a variation of $\alpha$ can
be modelled. From a `theoretical' point of view, the more convenient one
appears to be to interpret it as a variation in the speed of
light $c$ \cite{mof,abm,am,bim}, but other alternatives have been
explored \cite{beken}. It is also possible to
analyse the consequences of the variation of $\alpha$ in a more
phenomenological context, as was done in \cite{steen,us}.

On the observational level, the situation is at present somewhat
confusing---see \cite{varsh} for a brief summary. The best limit from
laboratory experiments (using atomic clocks) is \cite{prestage}
\begin{equation}
|{\dot \alpha}/\alpha|<3.7\times 10^{-14} yr^{-1}\, .
\label{labbound}
\end{equation}
Measurements of isotope ratios in the Oklo natural reactor provide the
strongest geophysical constraints \cite{damour},
\begin{equation}
|{\dot \alpha}/\alpha|<0.7\times 10^{-16} yr^{-1}\, ,
\label{oklobound}
\end{equation}
although
there are suggestions \cite{sisterna} that due to a number of nuclear physics
uncertainties and model dependencies a more realistic bound is
$|{\dot \alpha}/\alpha|<5\times 10^{-15} yr^{-1}$. Note that these
measurements effectively probe timescales corresponding to a cosmological
redshift of about $z\sim0.1$ (compare with astrophysical measurements below).

Three kinds of astrophysical tests have been used. Firstly, big bang
nucleosynthesis \cite{bbn}
can in principle provide rather strong constraints at very high redshifts, but
it has a strong drawback in that one is always forced to make
an assumption on how the neutron to proton mass difference depends on $\alpha$.
This is needed to estimate the effect of a varying $\alpha$ on the ${}^4He$
abundance. The abundances of the other light elements depend much less
strongly on this assumption, but on the other hand these abundances are
much less well known observationally. Hence one can only find the relatively
weak bound 
\begin{equation}
|{\Delta \alpha}/\alpha|<2\times 10^{-2}\, ,\qquad z\sim10^9-10^{10}.
\label{nuclbound}
\end{equation}

Secondly, observations of
the fine splitting of quasar doublet absorption lines
probe smaller redshifts, but should be much
more reliable. Unfortunately, the two groups which have been actively
studying this topic report different results. Webb and
collaborators \cite{webb} were the first to report a positive
result,
\begin{equation}
\Delta\alpha/\alpha=(-1.9\pm0.5)\times 10^{-5}\,,\qquad z\sim1.0-1.6
\label{webbpub}
\end{equation}
Note that
this means that $\alpha$ was {\em smaller} in the past. Recently the same
group reports two more (as yet unpublished) positive results \cite{webbnew},
$\Delta\alpha/\alpha=(-0.75\pm0.23)\times 10^{-5}$ for
redshifts $z\sim0.6-1.6$ and
$\Delta\alpha/\alpha=(-0.74\pm0.28)\times 10^{-5}$ for
redshifts $z\sim1.6-2.6$. On the other hand, Varshalovich and
collaborators \cite{varsh} report only a null result,
\begin{equation}
\Delta\alpha/\alpha=(-4.6\pm4.3\pm1.4)\times 10^{-5}\,,
\qquad z\sim2-4\,;
\label{varshbound1}
\end{equation}
the first error bar corresponds to the statistical error while the second
is the systematic one. This
corresponds to the bound
\begin{equation}
|{\dot \alpha}/\alpha|<1.4\times 10^{-14} yr^{-1}
\label{varshbound2}
\end{equation}
over a timescale of
about $10^{10}$ years. It should be emphasised that the observational
techniques used by both groups have significant differences, and it is
presently not clear how the two compare when it comes to eliminating
possible sources of systematic error.

Finally, a third option is the cosmic microwave background (CMB) \cite{steen}.
This probes intermediate redshifts, but has the significant advantage that
one has (or will soon have) highly accurate data.

The reason why the Cosmic Microwave Background is a good probe of
variations of the fine-structure constant is that these alter the
ionisation history of the universe \cite{steen,us}.
The dominant effect is a change
in the redshift of recombination, due to a shift in the energy levels
(and, in particular, the binding energy) of Hydrogen. The Thomson
scattering cross-section is also changed for all particles, being
proportional to $\alpha^2$. A smaller effect
(which has so far been neglected) is expected to come from a change in
the Helium abundance.

As is well known, CMB fluctuations are typically described in terms
of spherical harmonics,
\begin{equation}
T(\theta,\phi)=\sum_{\ell m}a_{\ell m}Y_{\ell m}(\theta,\phi)\, ,
\label{cmbsph}
\end{equation}
from whose coefficients one defines
\begin{equation}
C_\ell=<|a_{\ell m}|^2>\, .
\label{cmbcl}
\end{equation}

Increasing $\alpha$ increases the redshift
of last-scattering, which corresponds to a smaller sound horizon. Since the
position of the first Doppler peak (which we shall denote as $\ell_{peak}$)
is inversely proportional to the sound horizon at last scattering, we see
that increasing $\alpha$ will produce a larger $\ell_{peak}$ \cite{us}.
This larger redshift of last scattering also has the additional effect
of producing a larger early ISW effect, and hence a larger amplitude
of the first Doppler peak \cite{steen}. Finally, an increase
in $\alpha$ decreases the high-$\ell$ diffusion damping (which is
essentially due to the finite thickness of the last-scattering surface),
and thus increases the power on very small scales. 

The authors of \cite{steen} provide an analysis of these effects and
conclude that future CMB experiments should be able to provide
constraints on a varying $\alpha$ at the recombination epoch (that is, at
redshifts $z\sim1000$) at the level of
\begin{equation}
|{\dot \alpha}/\alpha|<7\times 10^{-13} yr^{-1}\, ,
\label{cmbbound1}
\end{equation}
or equivalently
\begin{equation}
|\alpha^{-1}d\alpha/dz|<9\times 10^{-5}\, ,
\label{cmbbound1a}
\end{equation}
which seems to indicate that these constraints can only become competitive
in the near future.

Here we analyse these effects for the BOOMERanG \cite{boom,lange} and MAXIMA \cite{maxima,balbi} data. We briefly
review the method and then discuss the results in the next section. We find
that this data tends to prefer a value of $\alpha$ that was lower in the past.
However, we strongly emphasise that there are interesting and so far
unnoticed degeneracies in the physics of the problem which imply that
this method of determining the fine-structure constant can only produce
strong constraints if other cosmological parameters are independently known.
We will comment on this point in section III.

While this paper was being finalised, another preprint appeared \cite{oth},
containing an independent analysis of the same data. It should be noticed
that there are some significant differences in the two analysis
procedures, as well as in the results,
which we will point out along the way. In the cases where a direct comparison
is possible, our work confirms their results, while
in the other cases we provide some physical motivation for the differences.

\section{Data analysis}
\label{data} 

We perform a likelihood analysis of the recently released BOOMERanG \cite{boom}
and MAXIMA \cite{maxima} data, allowing for the possibility of
a time-varying fine-structure
constant. The method used follows the procedure described in \cite{me98,me99}.
The angular power spectrum $C_l$ was obtained using a modified CMBFAST algorithm which allows a varying $\alpha$ parameter. We have changed
the subroutine RECFAST \cite{recfast} according
to the extensive description given in \cite{steen}.

We vary the power spectrum normalisation $C_2$ within
the $95 \%$ limits for the COBE 4-year data \cite{bennet}
The space of model parameters spans
\begin{equation}
\Omega_0=(0.1-1.0)\, ,
\label{varomega0}
\end{equation}
\begin{equation}
\Omega_{b}h^{2}=(0.01-0.028)\, ,
\label{varomegabh2}
\end{equation}
\begin{equation}
H_0=(50-80)\, ,
\label{varh0}
\end{equation}
\begin{equation}
\alpha / \alpha_0=(0.9-1.1)\, ,
\label{varalph}
\end{equation}
and the $C_l$ normalisation
\begin{equation}
bias=C_2/C_{2,COBE}=(0.83-1.16)\, .
\label{varbias}
\end{equation}
Note that $\alpha_0$ is the value of the fine structure constant today.
The basic grid of models was obtained considering parameter step sizes of 0.1 for 
$\Omega_0$; 0.003 for $\Omega_{b}h^{2}$; 5 for $H_0$; 0.01 for $\alpha / \alpha_0$
and finally 0.01 for the $bias$.
In order to compute the maxima and the confidence intervals for the 1-dim marginalised distributions we have
increased the grid resolution of each of the model parameters 
using interpolation procedures.

All our models have $\Omega_{total}=1$ and no tilt. We point out that
this is in agreement \cite{jaffe}
with the best-fit model for the $\Omega_{total}=1$ case for the
combined analysis of the BOOMERanG and MAXIMA data. Somewhat surprisingly,
the authors of \cite{oth} seem to find that the same data prefer tilted
models even in the `standard' case, that is
without considering a varying fine-structure constant. This will obviously
affect their results, as we will discuss below.

We also emphasise that $\alpha$ may or may not have had a different value
at the nucleosynthesis epoch, which in particular will affect the
value of $\Omega_{b}h^{2}$. Since in the present paper we do not
treat this effect, the correct approach \cite{am}
is to accept the observationally
determined value. This is the reason why we ignore excessively high values
of $\Omega_{b}h^{2}$ which seem to be required by some analyses of the
CMB data \cite{lange,jaffe}.
  
For each of these $C_l$ spectra we compute the flat band power estimates of the
CMB anisotropies obtaining a simulated observation for each data point. These
estimates cover the multi-pole range sampled by both BOOMERanG and MAXIMA
experiments. The values of $\Omega_b$ are obtained as {\em function}
of $H_0$ in
order to satisfy the range for $\Omega_{b}h^{2}$ defined above.

\subsection{The full dataset}

Interesting conclusions can be drawn by comparing the nominal calibration
case for both experiments with the Likelihood obtained after marginalising over the calibration
uncertainties of BOOMERanG ($20\%$) and MAXIMA ($8 \%$).
As mentioned in
\cite{white} the fitting to a lower second Doppler peak can be achieved either
by increasing the value of $\Omega_{b}h^{2}$ or decreasing the value of
$\alpha / \alpha_0$. Note also that there is an additional constraint
on $\alpha/\alpha_0$ coming from the position of the main acoustic
peak \cite{us}. We shall comment on the relative importance of the two
constraints below.

\begin{figure}
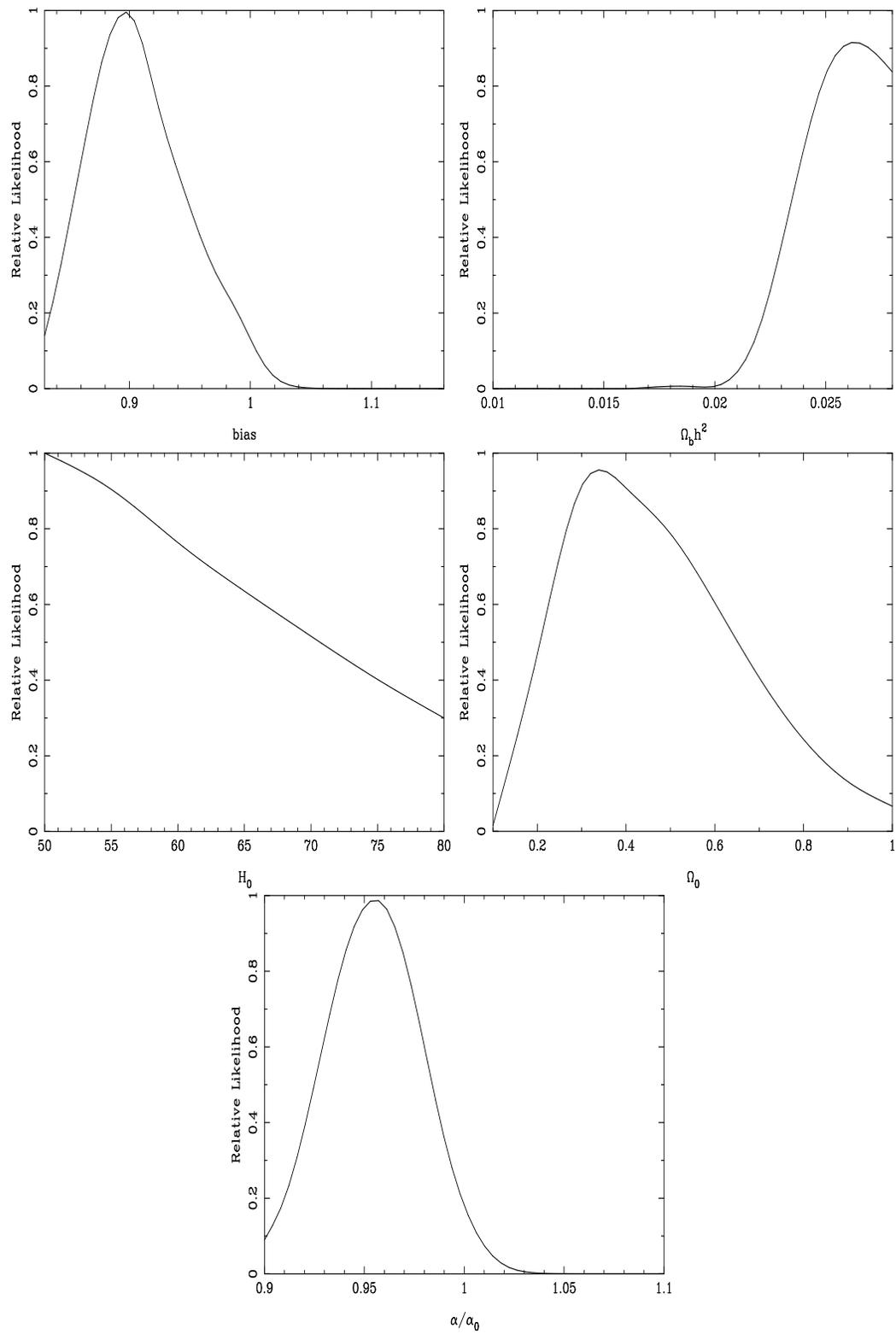

\hspace*{.5in}
\vbox{
\hbox{
\psfig{file=fig1a.ps,width=2.7in,height=2.7in,angle=270}
\psfig{file=fig1b.ps,width=2.7in,height=2.7in,angle=270}}
\hbox{
\psfig{file=fig1c.ps,width=2.7in,height=2.7in,angle=270}
\psfig{file=fig1d.ps,width=2.7in,height=2.7in,angle=270}}
\hspace*{1.3in}
\hbox{
\psfig{file=fig1e.ps,width=2.7in,height=2.7in,angle=270}}
}
\caption{Marginal distributions for each model parameter for 
the nominal calibration case.  
\label{nom_m}}
\end{figure}

For the nominal calibration case we obtain a best fit model with
\begin{equation}
H_0=55,\quad \Omega_{b}h^{2}=0.025,\quad \Omega_0=0.5\, ,
\label{fitcalnom1}
\end{equation}
\begin{equation}
\alpha / \alpha_0=0.94,\quad bias=0.87,\quad \chi^{2}=20.38\, .
\label{fitcalnom2}
\end{equation}
In Fig~\ref{nom_m} we plot the marginalised distributions for all
the parameters.

For this case the maxima and confidence intervals for the
marginalised distributions are as follows (all $1\sigma$):
$\alpha / \alpha_0=0.96^{+0.02}_{-0.03}$,
$H_0=50^{+16.02}_{-0.00}$, $\Omega_{b}h^{2}=0.026^{+0.002}_{-0.001}$,
$\Omega_0=0.3^{+0.3}_{-0.1}$; $bias=0.9 \pm 0.04$.

\begin{figure}
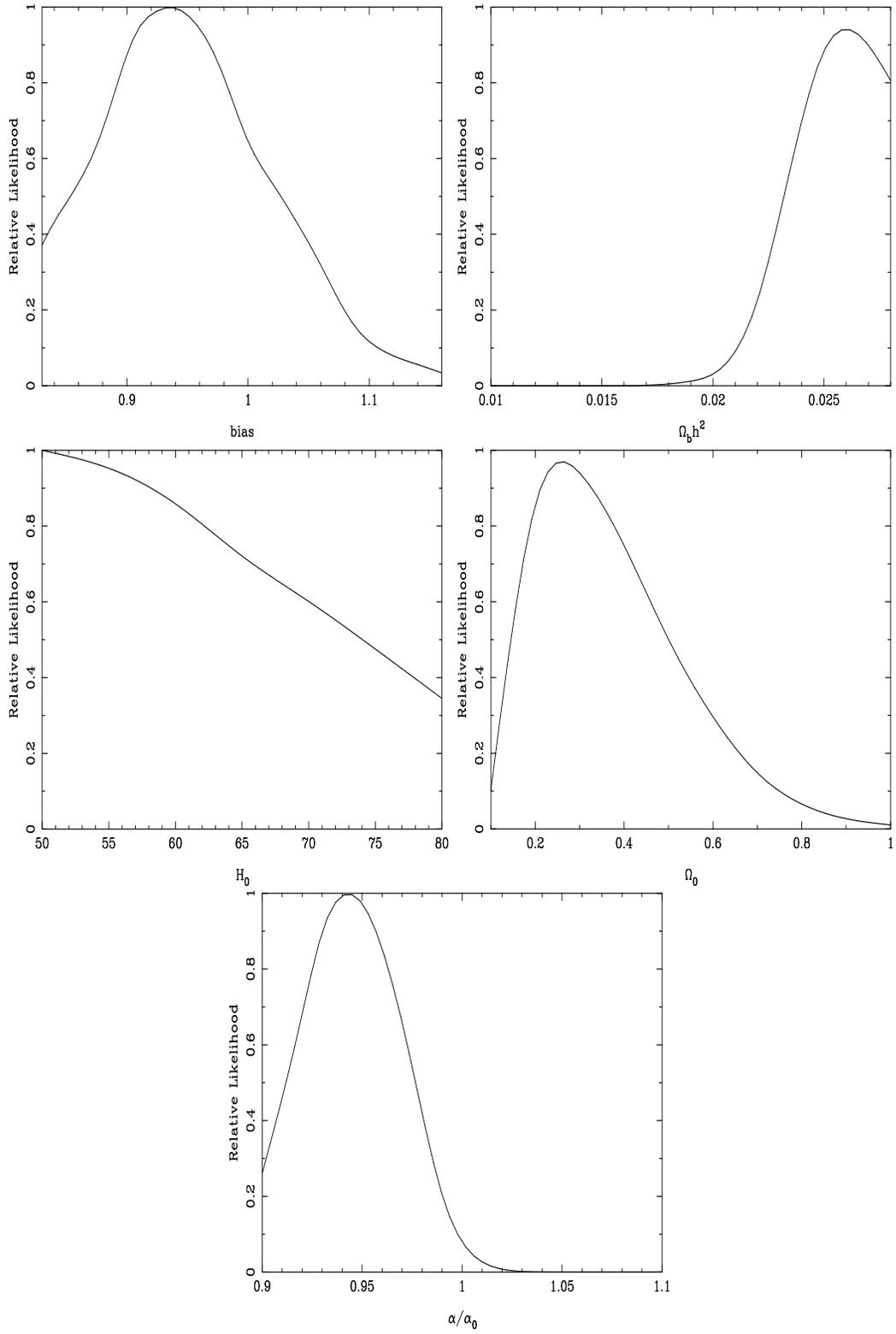

\hspace*{.5in}
\vbox{
\hbox{
\psfig{file=fig2a.ps,width=2.7in,height=2.7in,angle=270}
\psfig{file=fig2b.ps,width=2.7in,height=2.7in,angle=270}}
\hbox{
\psfig{file=fig2c.ps,width=2.7in,height=2.7in,angle=270}
\psfig{file=fig2d.ps,width=2.7in,height=2.7in,angle=270}}
\hspace*{1.3in}
\hbox{
\psfig{file=fig2e.ps,width=2.7in,height=2.7in,angle=270}}
}
\caption{Marginal distributions for each model parameter for the
calibration marginalised distribution.  
\label{cal_m}}
\end{figure}

We then marginalised the 5-dim Likelihood function over the calibration
uncertainties assuming a Gaussian prior. In this case
we obtain a best fit model with
\begin{equation}
H_0=55,\quad \Omega_{b}h^{2}=0.025,\quad \Omega_0=0.4\, ;
\label{fitcalmar1}
\end{equation}
\begin{equation}
\alpha / \alpha_0=0.93,\quad bias=0.91,\quad \chi^{2}=15.55\, .
\label{fitcalmar2}
\end{equation}
In Fig~\ref{cal_m} we plot the marginalised distributions for all the parameters for the distribution marginalised over the calibration errors. Comparing
with Fig~\ref{nom_m} we notice that in the case of the
marginalised distribution the distributions for $\Omega_0$ and
$\alpha/\alpha_0$ are slightly shifted towards lower values, while those of
$\Omega_{b}h^{2}$ and $H_0$ are not significantly affected.

The maxima and confidence intervals for the marginalised distributions are as follows
(again, all are $1\sigma$): 
$\alpha / \alpha_0=0.94^{+0.03}_{-0.02}$,
$H_0=50^{+16.62}_{-0.0}$; $\Omega_{b}h^{2}=0.026 \pm 0.002$,
$\Omega_0=0.3^{+0.17}_{-0.14}$, $bias=0.94 \pm 0.07$.

If we consider the best calibration case assuming a uniform
prior we get the same best fit set of parameter values 
as for the calibration marginalised likelihood, apart from the bias (which
now has the value $bias=0.84$).
The best calibration factor is  1.0 (nominal) for BOOMERanG
and 0.92 (ratio with respect to the nominal case)
for MAXIMA (this corresponds to $\chi^{2}=15.73$).
This is just telling us that if we keep BOOMERanG at the nominal calibration
case and lower the height of the MAXIMA data points we force the
normalisation of the models to decrease.
 
If instead we consider the best calibration case assuming a Gaussian
prior we get a best fit with a calibration factor of 1.1 for BOOMERanG
and 1.0 (nominal) for MAXIMA ($\chi^{2}=15.75$ before weighting).
As should be expected we get a higher
best fit value for the models normalisation of $bias=0.92$. Meanwhile for
both cases we observe a decrease on the value of $\Omega_0$ from 0.5 to 0.4
and on the value of $\alpha/ \alpha_0$ from 0.94 to 0.93, when compared
with the nominal case.

Note that both pushing the BOOMERanG data up or pushing the MAXIMA data
down provide for a better overlap of the two data sets. Then the different
overall normalisations (in particular the height of the first Doppler
peak) account for the different values of the cosmological parameters.

\begin{figure}
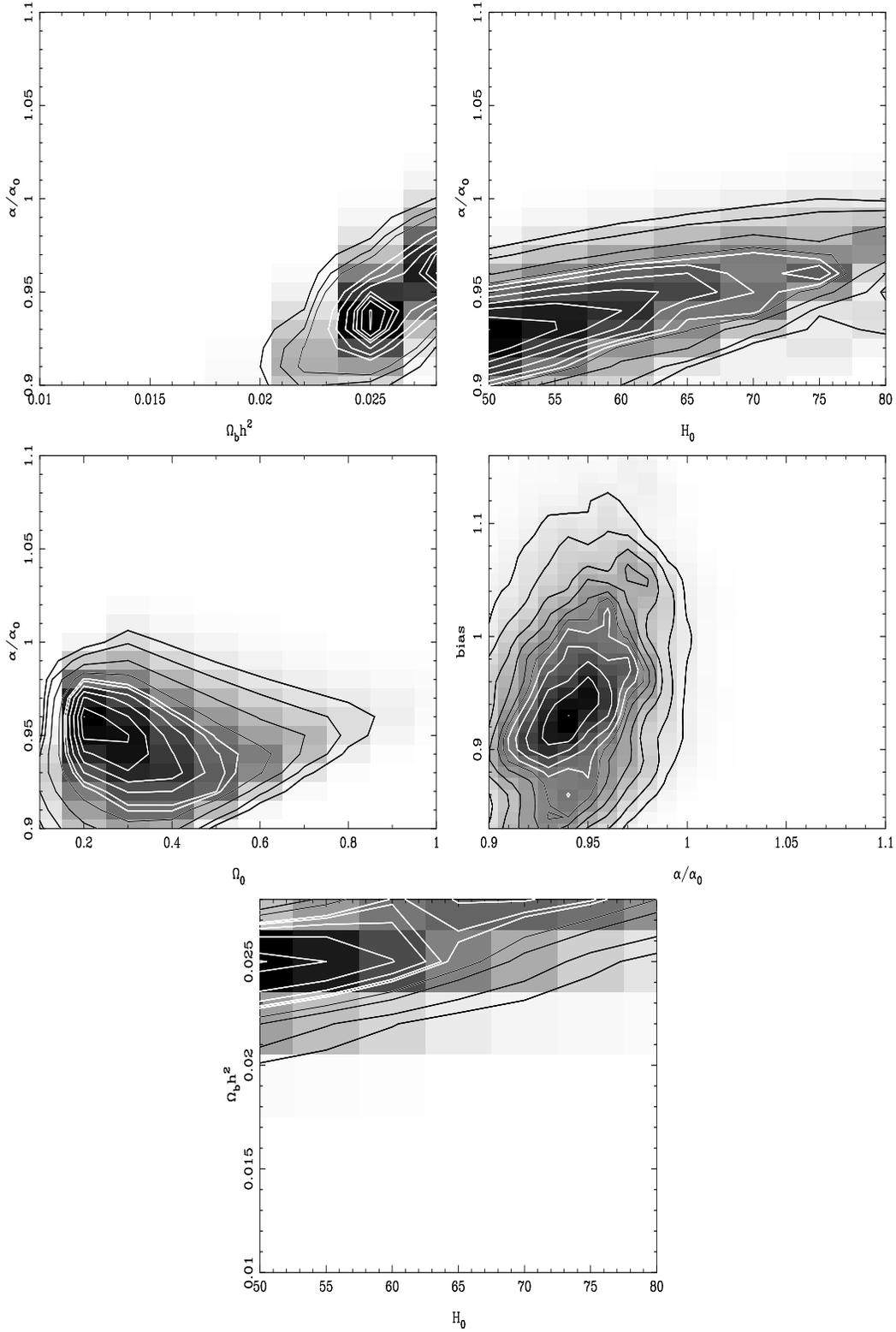

\hspace*{.5in}
\vbox{
\hbox{
\psfig{file=fig3a.ps,width=2.7in,height=2.7in,angle=270}
\psfig{file=fig3b.ps,width=2.7in,height=2.7in,angle=270}}
\hbox{
\psfig{file=fig3c.ps,width=2.7in,height=2.7in,angle=270}
\psfig{file=fig3d.ps,width=2.7in,height=2.7in,angle=270}}
\hspace*{1.3in}
\hbox{
\psfig{file=fig3e.ps,width=2.7in,height=2.7in,angle=270}}
}
\caption{Confidence contours for the 2-dim distribution obtained after
marginalising over the remaining three parameters (for the calibration
marginalised distribution). Contours are at 10,20,30,...90 and
95 per cent confidence levels.
\label{2dim_cal}}
\end{figure}

\begin{figure}
\hspace*{1.5in}
\vbox{
\psfig{file=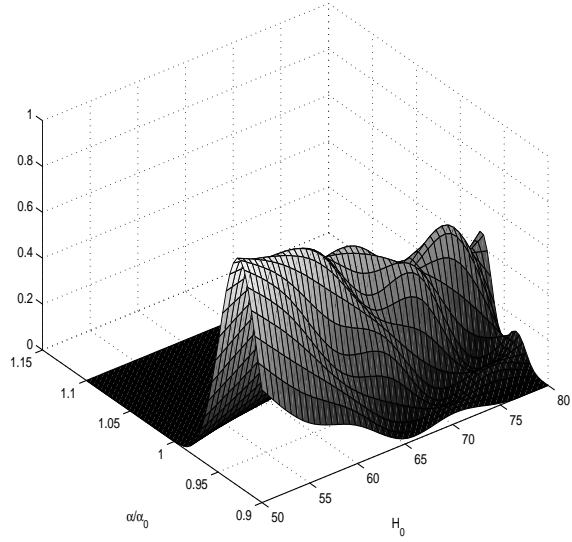,width=3in,height=3in}
\psfig{file=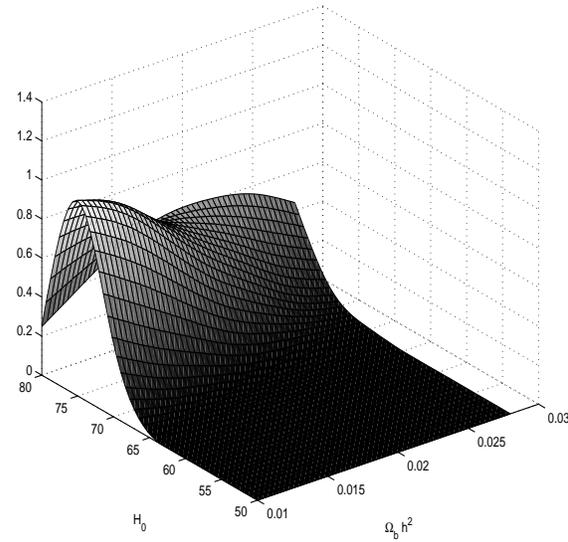,width=3in,height=3in}
}
\caption{The likelihood surface for $H_0$ and $\alpha / \alpha_0$ (top plot); for $\Omega_{b}h^{2}$ and $H_0$ (bottom plot), for the calibration marginalised distribution.
\label{3d}}
\end{figure}

\begin{figure}
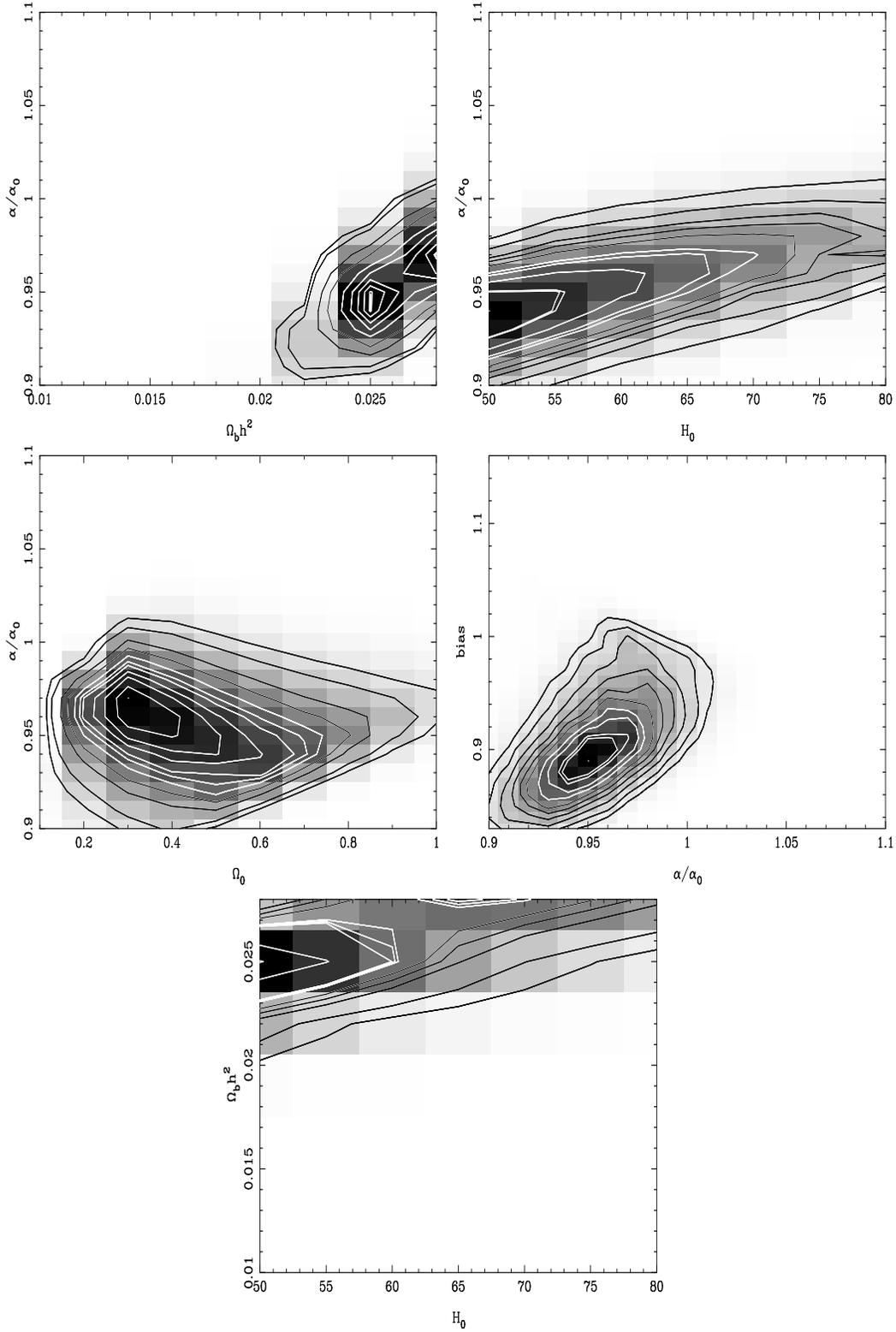

\hspace*{.5in}
\vbox{
\hbox{
\psfig{file=fig5a.ps,width=2.7in,height=2.7in,angle=270}
\psfig{file=fig5b.ps,width=2.7in,height=2.7in,angle=270}}
\hbox{
\psfig{file=fig5c.ps,width=2.7in,height=2.7in,angle=270}
\psfig{file=fig5d.ps,width=2.7in,height=2.7in,angle=270}}
\hspace*{1.3in}
\hbox{
\psfig{file=fig5e.ps,width=2.7in,height=2.7in,angle=270}}
}
\caption{Confidence contours for the 2-dim distribution obtained after
marginalising over the remaining three parameters (for the calibration
nominal case). Contours are at 10,20,30,...90 and
95 per cent confidence levels.
\label{2dim_nom}}
\end{figure}

In fig~\ref{2dim_cal} we plot the 2-dim Likelihood functions obtained
after marginalising over the remaining three parameters. In Fig~\ref{3d} 
we plot the likelihood surface for $H_0$ and $\alpha / \alpha_0$ as well as for $\Omega_{b}h^{2}$ and $H_0$. 
Similarly, Fig~\ref{2dim_nom} contains the corresponding likelihoods
for the nominal calibration case. This highlights the fact that there are
some non-trivial degeneracies in the problem \cite{hus}. We shall return
to this point below.

It is of interest to investigate the case where no variation of the fine
structure constant is allowed.
For that purpose we considered the conditional distribution
for $\alpha / \alpha_0=1.0$ to obtain a best fit model with
\begin{equation}
H_0=75,\quad \Omega_{b}h^{2}=0.028,\quad \Omega_0=0.3\, ;
\label{fitnoalpha1}
\end{equation}
\begin{equation}
bias=1.0,\quad \chi^2=17.49\, .
\label{fitnoalpha2}
\end{equation}

\begin{figure}
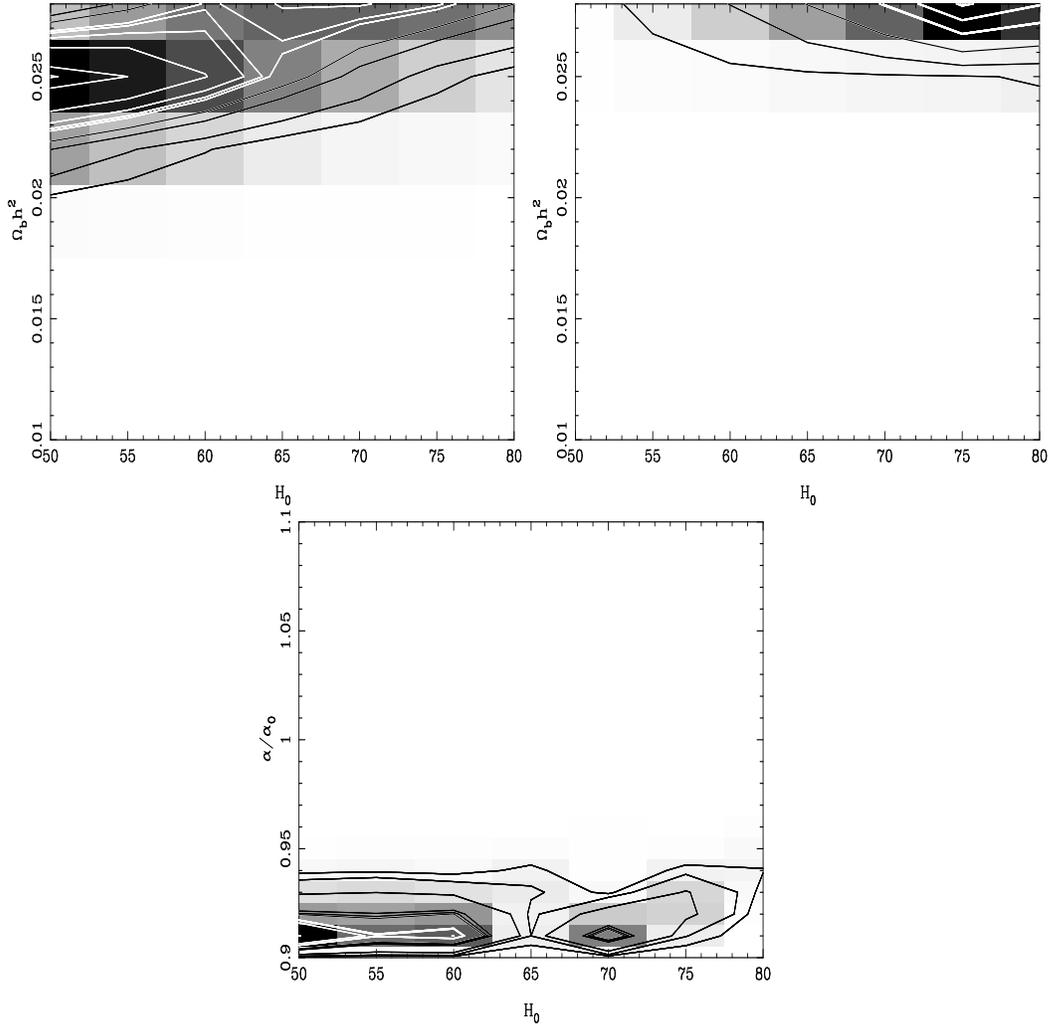

\hspace*{.5in}
\vbox{
\hbox{
\psfig{file=fig6a.ps,width=2.7in,height=2.7in,angle=270}
\psfig{file=fig6b.ps,width=2.7in,height=2.7in,angle=270}}
\hspace*{1.3in}
\psfig{file=fig6c.ps,width=2.7in,height=2.7in,angle=270}
}
\caption{Confidence contours for the 2-dim distribution of
($H_0$, $\Omega_{b}h^{2}$); marginalised over the remaining parameters
(left hand side plot) ; conditional distribution for $\alpha/ \alpha_0=1$
and marginalised over the remaining parameters (right hand side plot).
Bottom plot: Confidence contours for the 2-dim distribution of ($\alpha / \alpha_0$,$H_0$) conditional
to $\Omega_{b}h^{2}=0.019$ and marginalised over the remaining parameters
(for the calibration corrected likelihood). Contours are at 10,20,30,...90 and
95 per cent confidence levels.
\label{2dimcal_c}}
\end{figure}

In Fig~\ref{2dimcal_c} we plot this distribution marginalised
over $\Omega_0$ and the bias.
Increasing the value of $\alpha$ seems to force a higher best fit value
of $H_0$ and of $\Omega_{b}h^{2}$ and a lower value
of $\Omega_0$ with a best fit COBE normalised model. Again, this is
consistent with \cite{jaffe} (which also find a tilt $n_s=0.99$, while
\cite{oth} find $n_s=0.92$).

If we instead condition our distribution to a value
of $\Omega_{b}h^{2}=0.019$ we get a best fit model with
\begin{equation}
H_0=50,\quad  \Omega_0=0.3,\quad \alpha / \alpha_0=0.9\, ;
\label{fitnucleos1}
\end{equation}
\begin{equation}
bias=0.89,\quad \chi^{2}=18.23\, .
\label{fitnucleos2}
\end{equation}
This result emphasises the rather obvious point that reducing the value
of $\Omega_{b}h^{2}$ requires a lower value of $\alpha / \alpha_0$
to account for a low second acoustic peak.

Once more,
we emphasise that the values of $\Omega_b$ are obtained as function
of $H_0$ in order to satisfy the range for $\Omega_{b}h^{2}$ defined above.
This means that we should expect to observe some correlation when plotting
$H_0$ against $\Omega_{b}h^{2}$. This is indeed confirmed by
Fig~\ref{2dim_cal}.

Therefore we have so far confirmed the fact that to fit a low second Doppler
peak we need a high baryonic content \cite{hus,white} and a lower fine
structure constant in the past.
However, an important question still remains: what's the weight of the
second acoustic peak relative to the main peak in drawing the above
conclusions? We recall that in \cite{us} it was shown that the position of
the first Doppler peak can by itself provide a constraint on $\alpha$.
Can it happen that the main acoustic peak is still a heavy factor
in determining the above best fit parameters?

\subsection{The first Doppler peak}

In order to answer this important question, we considered the set of data
points sampling the multi-pole region to up $l\sim 400$. Hence this data set
consists now on the first 8 Boomerang and the first 5 Maxima data points with the nominal calibration ($BM_{dp}$). We then repeat the likelihood analysis for this new data set. 

We obtain a  best fit model with 
\begin{equation}
H_0=50,\quad \Omega_{b}h^{2}=0.019,\quad \Omega_0=0.4\, ,
\label{fitcalnom1a}
\end{equation}
\begin{equation}
\alpha / \alpha_0=0.9,\quad bias=0.87,\quad \chi^{2}=14.91\, .
\label{fitcalnom2a}
\end{equation}

This is rather encouraging, particularly because the best-fit value for
$\Omega_{b}h^{2}$ is precisely the one found by observations. However, 
the maxima and confidence intervals for the marginalised distributions
are as follows (again, all are $1\sigma$): 
$\alpha / \alpha_0=0.97 \pm 0.04$,
$H_0=71.4^{+6.1}_{-13.5}$; $\Omega_{b}h^{2}=0.024 \pm 0.003$,
$\Omega_0=0.4^{+0.37}_{-0.21}$, $bias=1.0^{+0.05}_{-0.08}$.

We conclude that most of the best fit model parameters do not lay within
the $1 \sigma$ range around the maximum of the marginalised distribution
for the corresponding parameter. Therefore the 5-dim likelihood must have
a narrow peak around this best model with an enlarged surface around
the remaining values which height is not significantly smaller then the
absolute peak.

\begin{figure}
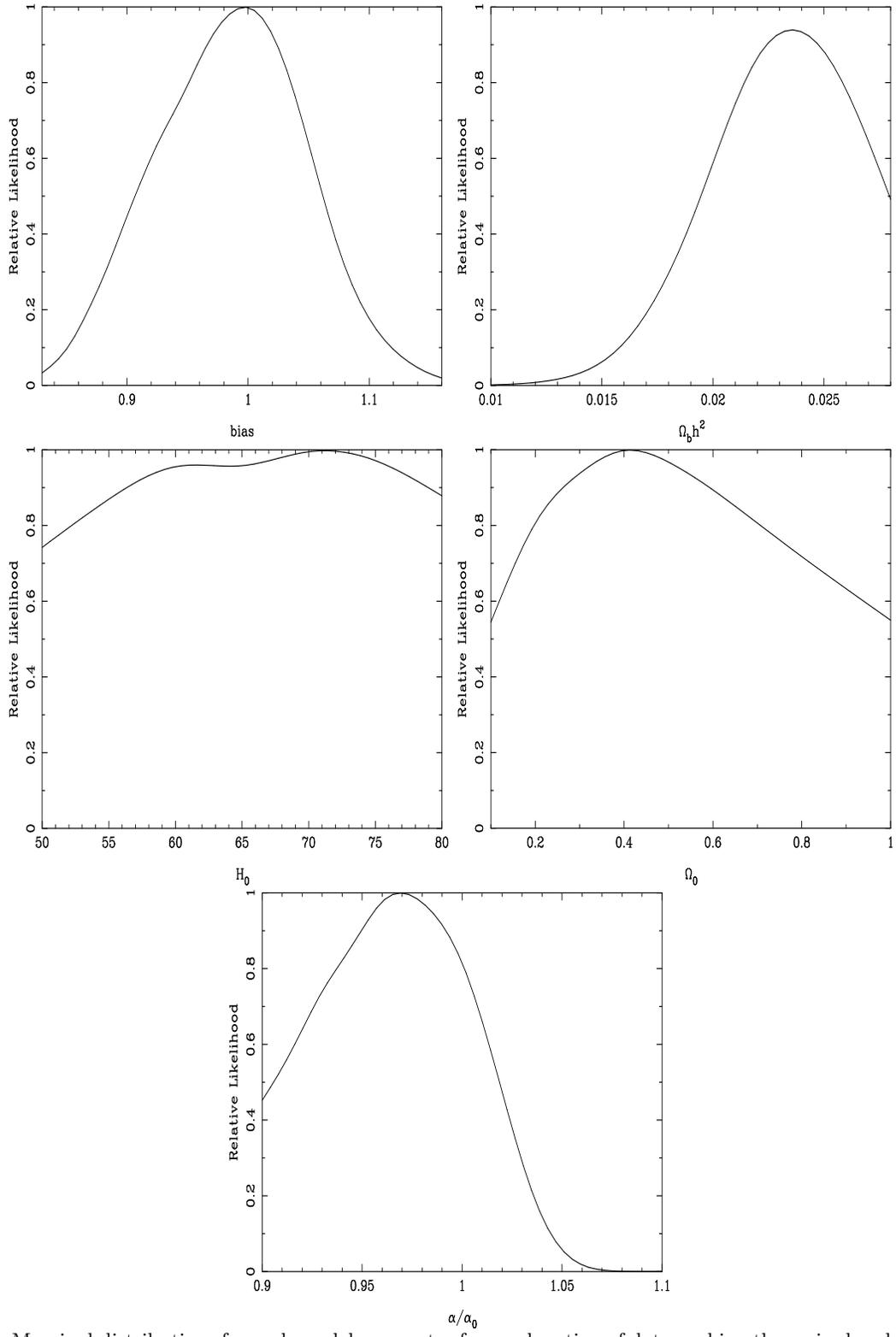

\hspace*{.5in}
\vbox{
\hbox{
\psfig{file=fig7a.ps,width=2.7in,height=2.7in,angle=270}
\psfig{file=fig7b.ps,width=2.7in,height=2.7in,angle=270}}
\hbox{
\psfig{file=fig7c.ps,width=2.7in,height=2.7in,angle=270}
\psfig{file=fig7d.ps,width=2.7in,height=2.7in,angle=270}}
\hspace*{1.3in}
\hbox{
\psfig{file=fig7e.ps,width=2.7in,height=2.7in,angle=270}}
}
\caption{Marginal distributions for each model parameter for a subsection of data probing 
the main doppler peak region (and nominal calibration case) ($BM_{dp}$).  
\label{dp_nom_m}}
\end{figure}

\begin{figure}
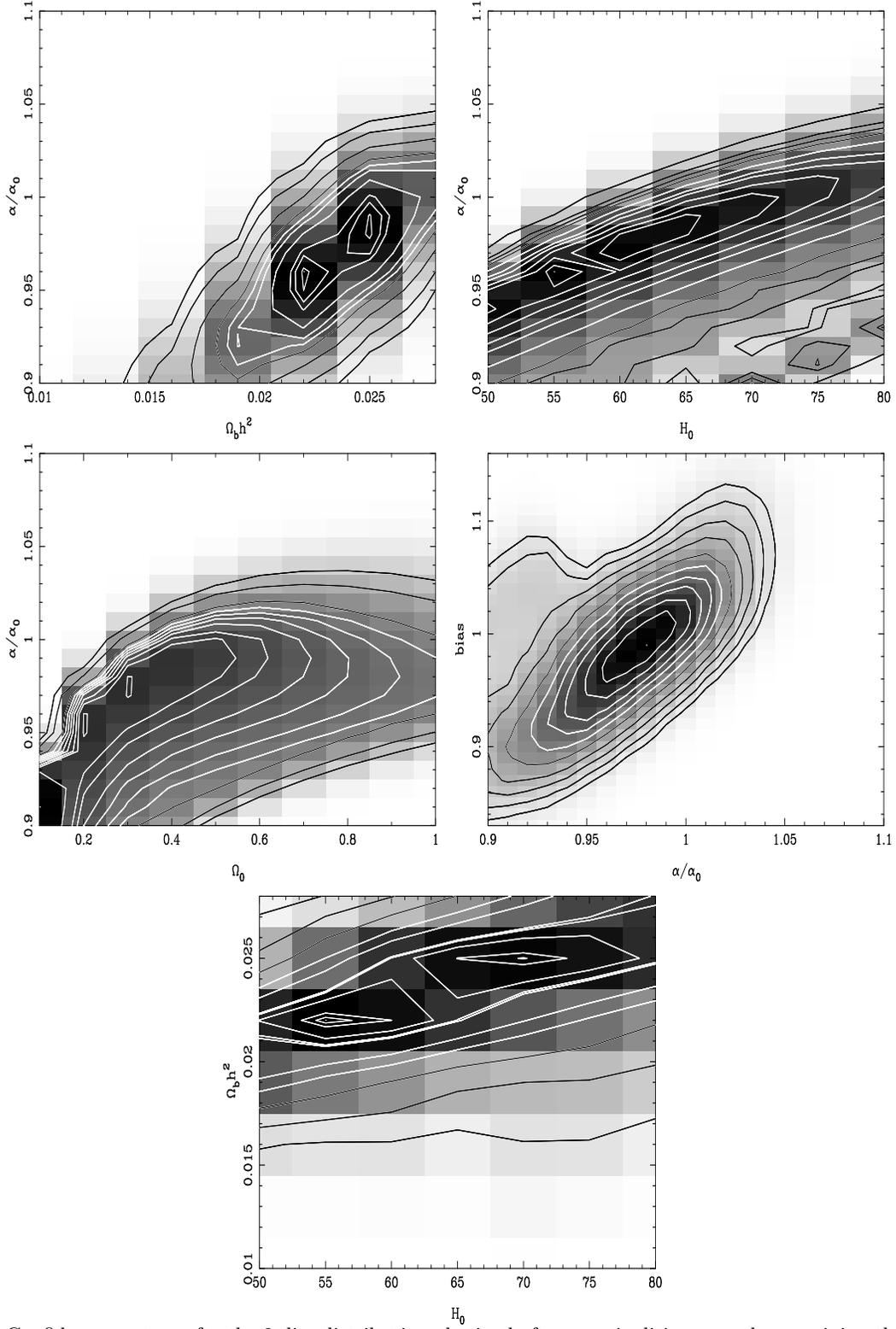

\hspace*{.5in}
\vbox{
\hbox{
\psfig{file=fig8a.ps,width=2.7in,height=2.7in,angle=270}
\psfig{file=fig8b.ps,width=2.7in,height=2.7in,angle=270}}
\hbox{
\psfig{file=fig8c.ps,width=2.7in,height=2.7in,angle=270}
\psfig{file=fig8d.ps,width=2.7in,height=2.7in,angle=270}}
\hspace*{1.3in}
\hbox{
\psfig{file=fig8e.ps,width=2.7in,height=2.7in,angle=270}}
}
\caption{Confidence contours for the 2-dim distribution obtained after
marginalising over the remaining three parameters for $BM_{dp}$. Contours are at 10,20,30,...90 and 95 per cent confidence levels.
\label{2dim_dp_nom}}
\end{figure}

In Fig~\ref{dp_nom_m} we plot the marginalised distributions for all
the parameters, while in fig~\ref{2dim_dp_nom} we plot the 2-dim
Likelihood functions obtained after marginalising over the remaining
three parameters.

Comparing Fig~\ref{dp_nom_m} with Fig~\ref{nom_m} we immediately notice
a number of extremely interesting points.
Firstly, even though the dataset for the first Doppler peak favours a
smaller $\alpha$ in the past, there is a non-negligible likelihood
for {\em larger} values as well. The inclusion of information from the
second Doppler peak all but eliminates this possibility (compare this with
Fig.~4 of \cite{oth}).
Secondly, the full dataset {\em increases} the preferred values of
$\Omega_{b}h^{2}$ (or more accurately, decreases the probability for
low values). And thirdly, data from the first Doppler peak alone is basically
insensitive to $H_0$ and $\Omega_0$ (recall that all our models have
$\Omega_{total}=1$), while the full dataset
tends to favour low values of $H_0$ and also narrows the distribution
for $\Omega_0$ around a value of $\Omega_0=0.3$ (and most notably
reduces the probability of lower values such as $\Omega_0=0.1$
which would be allowed by the reduced dataset).

This might 
explain the differences in the contour plots of the 2-dim distribution of ($\alpha/ \alpha_0$, $\Omega_0$) in Fig~\ref{2dim_dp_nom} and Fig~\ref{2dim_nom}; the plot in Fig~\ref{2dim_dp_nom} shows
a correlation between ($\alpha/ \alpha_0$, $\Omega_0$)
which disappears when including the other Doppler 
peaks. These 2-dim plots do also indicate correlations
between ($\alpha/ \alpha_0$, $H_0$); 
($\alpha/ \alpha_0$,$\Omega_{b}h^{2}$) and  ($\alpha/ \alpha_0$,bias)
which do exist in both 
situations.
  
Finally, we also consider the case with no variation of the fine
structure constant allowed for the reduced dataset.
We obtain a best fit model with
\begin{equation}
H_0=75,\quad \Omega_{b}h^{2}=0.025,\quad \Omega_0=0.5\, ;
\label{fitnoalpha1a}
\end{equation}
\begin{equation}
bias=1.02,\quad \chi^{2}=15.24\, .
\label{fitnoalpha2a}
\end{equation}

If now we condition our distribution to a value
of $\Omega_{b}h^{2}=0.019$ we get a best fit model with
\begin{equation}
H_0=50,\quad  \Omega_0=0.4,\quad \alpha / \alpha_0=0.9\, ;
\label{fitnucleos1a}
\end{equation}
\begin{equation}
bias=0.87,\quad \chi^{2}=14.91\, .
\label{fitnucleos2a}
\end{equation}

\begin{figure}
\hspace*{.5in}
\vbox{
\hbox{
\psfig{file=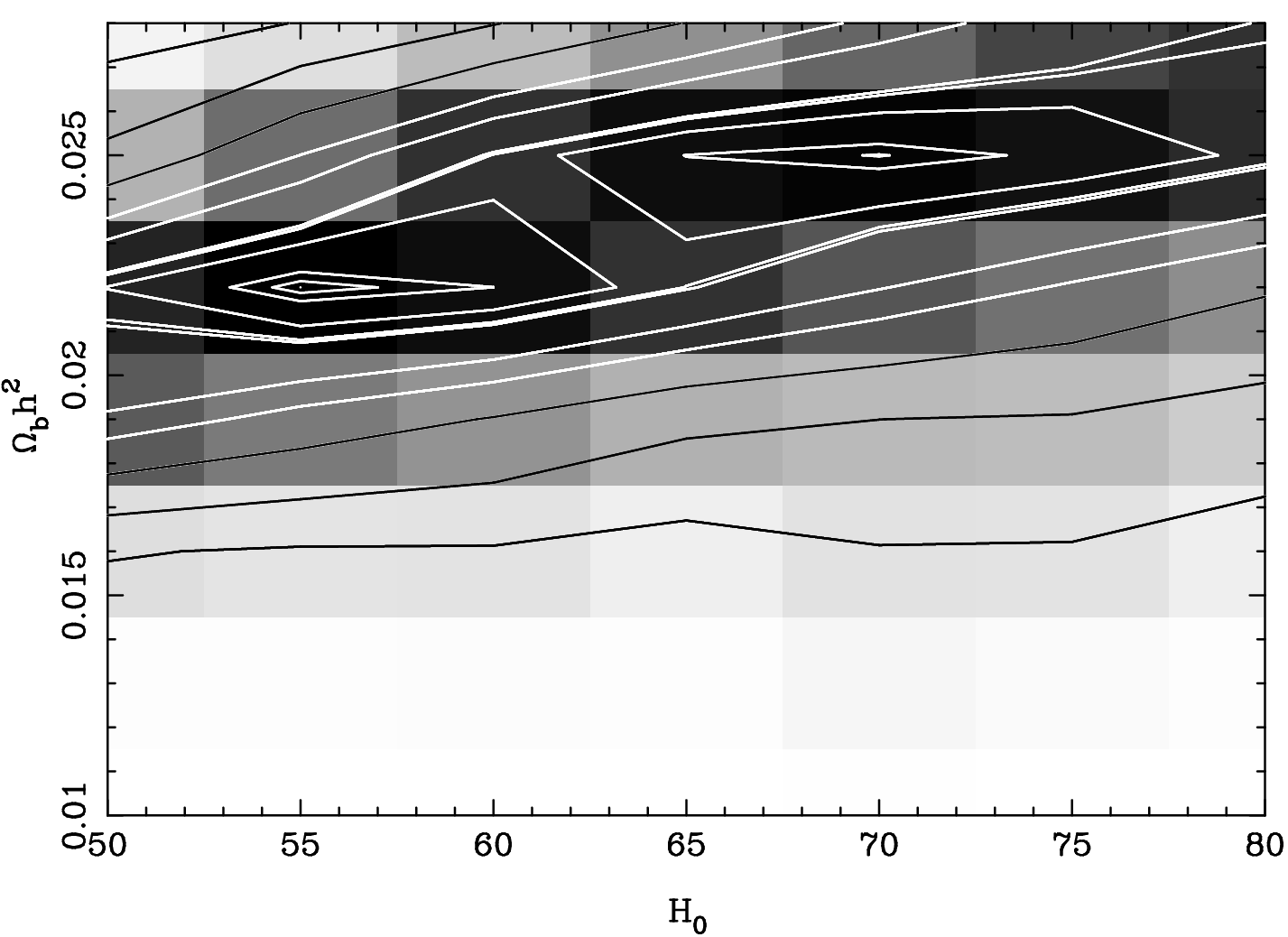,width=2.7in,height=2.7in,angle=270}
\psfig{file=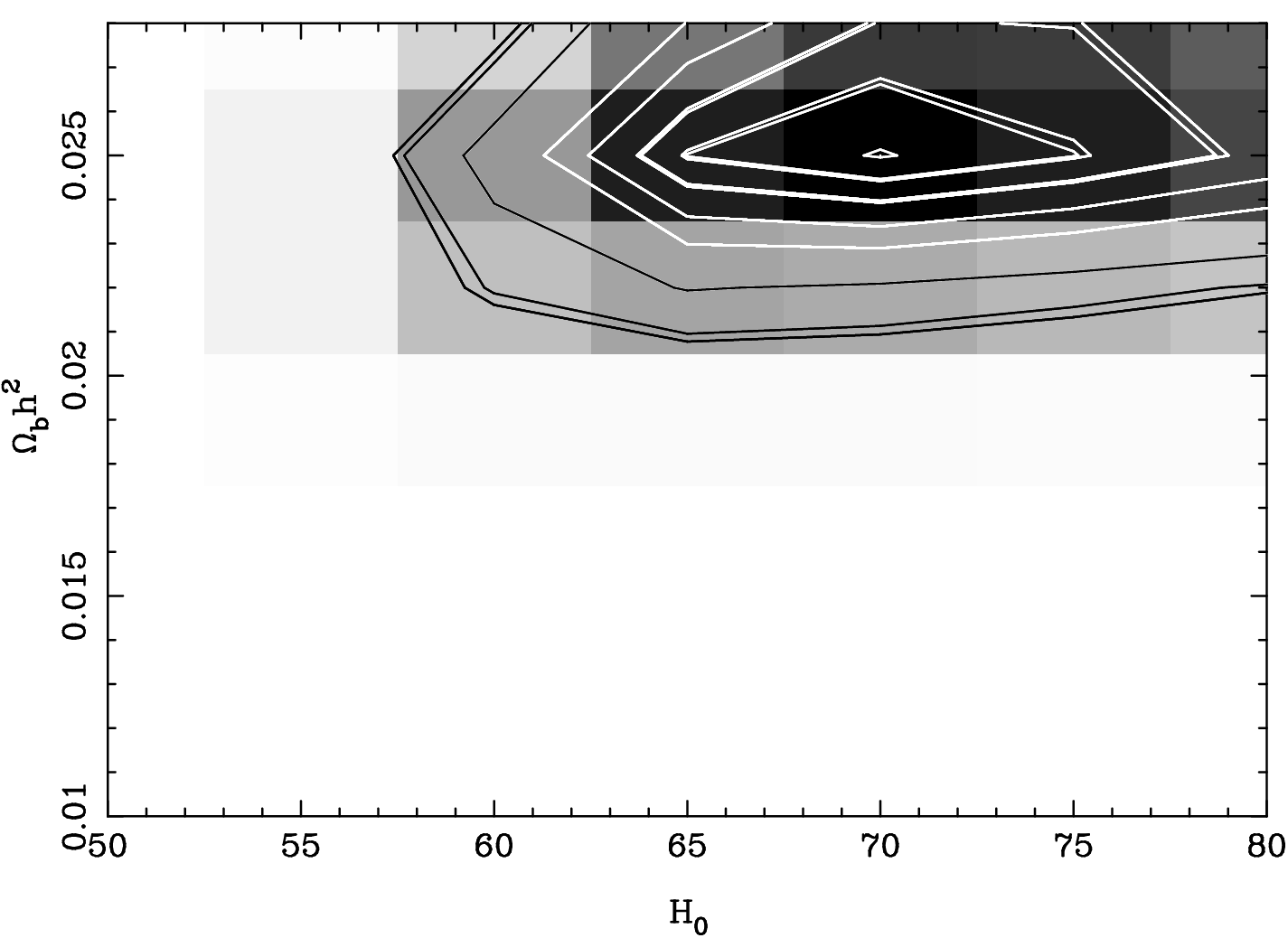,width=2.7in,height=2.7in,angle=270}}
\hspace*{1.3in}
\psfig{file=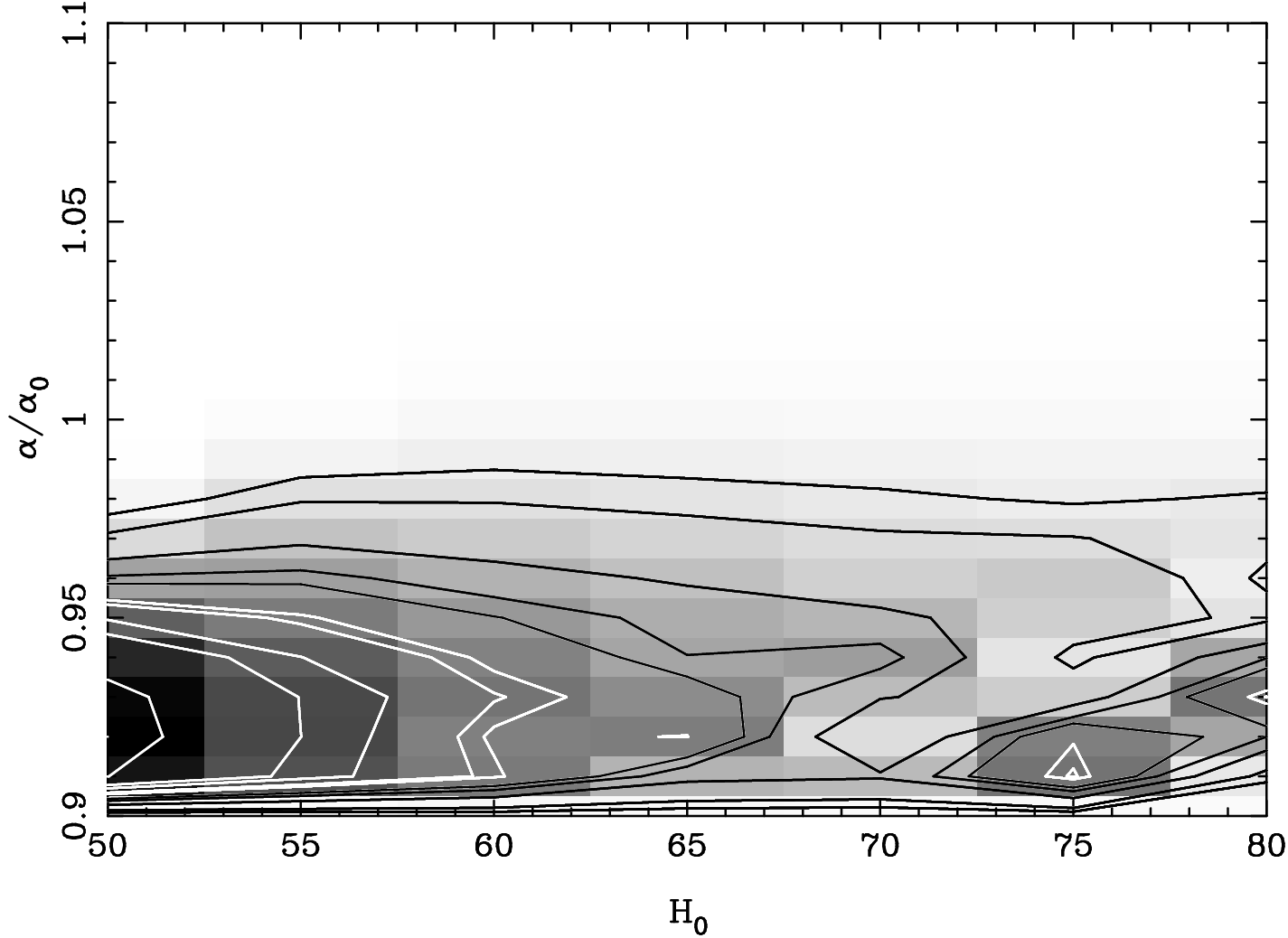,width=2.7in,height=2.7in,angle=270}
}
\caption{Confidence contours for the 2-dim distribution of
($H_0$, $\Omega_{b}h^{2}$); marginalised over the remaining parameters
(left hand side plot) ; conditional distribution for $\alpha/ \alpha_0=1$
and marginalised over the remaining parameters (right hand side plot).
Bottom plot: Confidence contours for the 2dim distribution of ($\alpha / \alpha_0$,$H_0$) conditional
to $\Omega_{b}h^{2}=0.019$ and marginalised over the remaining parameters
(for $BM_{dp}$). Contours are at 10,20,30,...90 and
95 per cent confidence levels.
\label{2dimdp_nom_c}}
\end{figure}

In Fig~\ref{2dimdp_nom_c} we plot these conditional distributions
marginalised over $\Omega_0$ and the bias.

\section{Discussion and conclusions}
\label{concl}

In this paper we have performed a likelihood analysis of
the combined BOOMERanG and MAXIMA datasets,
allowing for the possibility of a time-varying fine-structure
constant, for which there is further observational
evidence elsewhere \cite{webb,webbnew}. We have confirmed the intuitively
obvious expectation that this data prefers a value of $\alpha$ that
was smaller in the past by a few per cent.

However, we wish to emphasise that this is not the same as saying that the
CMB can readily provide and unambiguous measurement of the fine-structure
constant. As we hopefully made clear above, there are some interesting
degeneracies in the problem which imply that other cosmological parameters
could still fairly easily mimic a varying $\alpha$.
Hence this method of measurement of $\alpha$ is still far from
being `competitive', in the sense that statements about $\alpha$ will not
be possible until independent accurate determinations of
$\Omega_b h^2$ and $H_0$ (and possibly other parameters) are available.

We have also shown that the main reason behind the preferred lower value
of $\alpha$ in the present dataset still comes mainly from the
data points around the first Doppler peak. The main effect
of the high-$\ell$ data points is to increase the preferred
value for $\Omega_b h^2$ and eliminate the possibility of a larger
fine-structure constant in the past. A secondary (from this perspective)
effect of the small angular scale data is to tighten the constraints
on other parameters. Furthermore, we believe that this relative dominance
of the low-$\ell$ measurements will remain even in the post-MAP era.

\acknowledgements

We thank Bruce Bassett, Pedro Ferreira,
Jo\~ao Magueijo, Anupam Mazumdar and Paul Shellard
for useful discussions and comments.
C.M. and G.R. are funded by FCT (Portugal) under
`Programa PRAXIS XXI', grant no. PRAXIS XXI/BPD/11769/97 and 
PRAXIS XXI/BPD/9990/96, respectively.
We thank Centro de Astrof{\' \i}sica 
da Universidade do Porto (CAUP) for the facilities provided.
GR also thanks the Dept. of Physics of the University of Oxford for support 
and hospitality during the progression of this work.

\end{document}